\documentclass[twocolumn,showpacs,amsmath,amstex,amssymb,mathfonts,prb]{revtex4-1}
\usepackage{graphicx,bm,units}

\begin{document}

\title{Effect of Strong Disorder in a 3-Dimensional Topological Insulator: Phase Diagram and Maps of the ${\bm Z}_2$ Invariant}

\author{Bryan Leung$^1$ and Emil Prodan$^2$}
\address{$^1$Center for Materials Theory, Department of Physics \& Astronomy, Rutgers University, Piscataway NJ 08854, USA \\
$^2$Department of Physics, Yeshiva University, New York, NY 10016, USA} 

\begin{abstract} We study the effect of strong disorder in a 3-dimensional topological insulators with time-reversal symmetry and broken inversion symmetry. Firstly, using level statistics analysis, we demonstrate the persistence of delocalized bulk states even at large disorder. The delocalized spectrum is seen to display the levitation and pair annihilation effect, indicating that the delocalized states continue to carry the ${\bm Z}_2$ invariant after the onset of disorder. Secondly, the ${\bm Z}_2$ invariant is computed via twisted boundary conditions using an efficient numerical algorithm. We demonstrate that the ${\bm Z}_2$ invariant remains quantized and non-fluctuating even after the spectral gap becomes filled with dense localized states. In fact, our results indicate that the ${\bm Z}_2$ invariant remains quantized until the mobility gap closes or until the Fermi level touches the mobility edges. Based on such data, we compute the phase diagram of the Bi$_2$Se$_3$ topological material as function of disorder strength and position of the Fermi level. 
 
\end{abstract}


\date{\today}

\maketitle

Topological insulators represent a new class of materials where the topology of the bulk electronic structure induces highly non-trivial effects.\cite{HALDANE:1988rh,Kane:2005np,Kane:2005zw,Bernevig:2006hl,Koenig:2007ko,Moore:2007ew,Fu:2007vs,Hsieh:2008vm} One such effect is the emergence of metallic states along the edges of planar (2D) topological insulating structures or on the surface of 3-dimensional (3D) topological insulators. These materials became a reality after a topological insulator from the class of Quantum Spin-Hall insulators and one from the class of strong 3D topological insulators with time reversal symmetry have been theoretically predicted and then engineered and characterized in laboratories.\cite{Bernevig:2006hl,Koenig:2007ko,Moore:2007ew,Fu:2007vs,Hsieh:2008vm} Since then, many additional topological materials have been discovered (the reader can find a survey of the field in Refs.~\onlinecite{Qi2010,ZHassanRevModPhys2010du,QiRMP2011tu,Hasan2010by}).

For topological insulators with time-reversal symmetry, it has been argued that the edge/surface states maintain their metallic character even in the presence of weak disorder,\cite{Kane:2005np,Kane:2005zw} due to the cancelation of the backscattering amplitudes. This robustness against disorder can be the key to many technological applications,\cite{Qi2010,ZHassanRevModPhys2010du,QiRMP2011tu} and because of that, a great deal of effort has been dedicated to understanding the behavior of the topological materials in the presence of disorder.\cite{Sheng:2006na,Onoda:2007xo,Essin:2007ij,Obuse:2007qo,Obuse:2008ff,LiPRL2009xi,GrothPRL2009xi,JiangPRB2009sf,Yamakage2010xr,Prodan2010ew,GuoPRB2010fu,GuoPRL2010vb,Prodan2011vy,ProdanJPhysA2011xk} One important question, which is still opened for 3D topological insulators, is if the robustness against disorder extends into the strong disorder regime, particularly into the regime where the insulating gap is filled with dense localized spectrum. The theoretical argument based on the time-reversal symmetry is perturbative and therefore it breaks down in this regime. As such, one must seek a new argument that combines topology (an index theorem) and symmetry and this has become an extremely active area of research.\cite{Loring2010vy,HastingsAnnPhys2011gy,ProdanJPhysA2011xk} While searching for such argument, it became apparent to us that a numerical exploration of the matter will be of great help. Notable efforts paralleling ours are the numerical computations of a newly defined Bott index in Ref.~\onlinecite{HastingsAnnPhys2011gy}, carried out for a disordered model of a 3D strong topological insulator. The explicit equivalence between the Bott index and the strong ${\bm Z}_2$ invariant remains to be established in the strong disorder regime considered here. We also want to mention the scattering approach for disordered topological insulators reported in Ref.~\onlinecite{FulgaPRB2012cg}.

Our discussion will be restricted to 3D topological insulators. In the presence of time reversal symmetry, the insulators follow a ${\bm Z}_2$ topological classification. The strong ${\bm Z}_2$ invariant that renders an insulator as either trivial or topological was computed in various ways, but in general the computations were quite demanding because they had to be carried out with special smooth gauges. The difficulty introduced by this requirement has been documented in our previous work.\cite{ProdanPRB2011vy} For example, the original expressions of the ${\bm Z}_2$ invariant,\cite{Kane:2005zw,Fu:2006ka,Fu:2007ti,Fu:2007vs} require special smooth gauges and were computed only for analytically solvable band models. These expressions have been reformulated in an almost gauge invariant fashion by Fukui and his collaborators.\cite{FukuiJPSJ2005gu,FukuiJPSJ2007ny} The method still requires a time-reversal adapted gauge at the boundary of half of the Brillouin zone, but nevertheless it became the method of choice when computing the ${\bm Z}_2$ invariant.\cite{Essin:2007ij,XiaPRL2010ni,FengPRB2010ry,SoluyanovPRB2011bt,FengPRL2010gu,WadaPRB2011fu} Still, an application of the method to the disordered case exists only in 2D.\cite{Essin:2007ij} The Chern-Simons integral of the quantized magneto-electric polarization also requires a globally smooth gauge.\cite{QiPRB2008ng} The difficulties introduced by this requirement were highlighted in Ref.~\onlinecite{CohPRB2011rt}, where the best effort to evaluate the Chern-Simons integral only led to a  value of 0.3, for a disorder-free topological case where the result should have been quantized to 1. To date, there is no successful direct evaluation of this Chern-Simons integral for clean tight-binding models.

The issue was recently reconsidered and gauge-independent formulations of the weak and strong ${\bm Z}_2$ invariants are now available.\cite{Ringel2010vo,SoluyanovPRB2011bt,Soluyanov2011gy,Yu2011re,ProdanPRB2011vy} Here we will follow Ref.~\onlinecite{ProdanPRB2011vy} and we will argue here that this new formulations bring certain numerical advantages which open the possibility of directly computing the ${\bm Z}_2$  invariants for systems with extremely large unit cells, particularly for disordered samples (as opposed to indirectly inferring the ${\bm Z}_2$ invariants from other type of calculations such as transport simulations of the surface states). We present a numerical analysis of the strong ${\bm Z}_2$ invariant for a system without inversion symmetry, computed in the weak and strong disorder regimes via the twisted boundary conditions technique combined with the new formulation of the invariant. The use of the twisted boundary conditions was advocated by Kane and Mele in their original discussion of the 2D ${\bm Z}_2$ invariant as an effective procedure for tackling the effect of disorder and electron-electron interaction.\cite{Kane:2005zw} Numerically, this method is equivalent to computing the ${\bm Z}_2$ invariant for a periodic system with a very large unit cell, leading to thousands of occupied energy bands. Finding smooth special gauges for such complex band structures is prohibitively difficult, which is why the ${\bm Z}_2$ invariant is notoriously difficult to compute for disordered 3D topological insulators (the parity analysis was accomplished in Ref.~\onlinecite{GuoPRB2010fu} for a system and disorder with inversion symmetry).

Working with a tight-binding model appropriate for the 3D topological material Bi$_2$Se$_3$, we provide compelling evidence that the strong ${\bm Z}_2$ invariant remains well defined and quantized even after the insulating gap becomes filled with dense localized spectrum. In fact, our various mappings of  the ${\bm Z}_2$ invariant indicate that the quantization holds as long as the Fermi level remains in the mobility gap. Furthermore, we use level statistics analysis to map the localized or extended character of the energy spectrum for a wide range of disordered strengths. We provide compelling evidence that there are bulk metallic states that persist even at very large disorder and we derive the phase diagram of the 3D model as function of Fermi level and disorder strength. The phase diagram consists of the strong topological phase, which is completely surrounded by a metallic phase, which is again surrounded by the trivial insulating phase. Computations of the ${\bm Z}_2$ invariant along paths that cross from the topological into the trivial insulating phase reveal strict quantization of the invariant to $\pm1$ values in the topological/trivial insulating phases, respectively, and strong fluctuations between $\pm 1$ inside the metallic phase.

The motivation behind the present study was two-fold. Firstly, there is no theory of the ${\bm Z}_2$ invariant for aperiodic systems (except for the trivial case when the Fermi level is in a spectral gap, i.e. a region that is void of any energy spectrum). For example, for the Chern and spin-Chern invariants we have theories that provide explicit and specific conditions, which can be written in one line, that tells us when these invariants take quantized values even if the Fermi level is not in a spectral gap.\cite{BELLISSARD:1994xj,ProdanJPhysA2011xk} Furthermore, we have real-space formulations of these two invariants,\cite{Prodan2010ew,ProdanJPhysA2011xk} which allows one to compute them in ``one shot" without involving twisted boundary conditions. Nothing like that exists for the strong ${\bm Z}_2$ invariant, despite some sustained efforts. This makes one to question that such a theory will ever be achieved, and in fact to question that the strong ${\bm Z}_2$ invariant does indeed remain quantized once the spectral gap is closed. Our numerical study provides the first direct evidence that the strong ${\bm Z}_2$ invariant behaves similarly with these other two invariants, which can be a strong motivation for people to continue searching for a theory of the strong ${\bm Z}_2$ invariant for aperiodic systems. 

Secondly, it is known that disorder can strongly deform the phase boundary of the topological state.\cite{LiPRL2009xi,GrothPRL2009xi,GuoPRL2010vb,Prodan2011vy} As such, it is highly desired to devise quantitative methods that can accurately pinpoint the extent of the topological phases. Previously, the strong topological phase was identified by probing the metallic character of the surface states via transport calculations in a long bar geometry.\cite{GuoPRB2010fu} For the special case when the system and the disorder have inversion symmetry, the topological phase was identified using the parities of the states.\cite{GuoPRB2010fu} Ideally, however, it will be to directly map the strong ${\bm Z}_2$ invariant and our study demonstrates that this is indeed possible for 3D materials. 

Lastly, we want to state explicitly that our numerical simulations probe un-charted territories. As discussed in the next sections, our method, and for that matter all the established methods, are easily seen to produce quantized and non-fluctuating ${\bm Z}_2$ values, if a spectral gap between the occupied and non-occupied levels remains open at all times while twisting of the boundary conditions. However, at strong disorder, the spectral gap not only closes but the levels can change their ordering when twisting the boundary conditions. Thus, levels that once were occupied become un-occupied and vice-versa. Moreover, different disorder configurations can no longer be connected adiabatically. While our numerical procedure can still be applied in these situations, the available theoretical arguments can no longer assure us that the output remains the same from one disorder configuration to another (this also applies to the Bott index of Refs.~\onlinecite{Loring2010vy,HastingsAnnPhys2011gy}).  Still, if the states near the Fermi level are localized, one expects the value of the invariant to remain un-affected by this phenomenon. This is exactly what we are trying to verify in this work.

\section{The twisted boundary conditions}

In the first half of the paper we will carry the discussion at a general level. We consider a generic 3D quantum lattice model with many quantum states $\xi$ per site ${\bm n}$. The Hilbert space ${\cal H}$ is spanned by $|{\bm n},\xi\rangle$ and the periodic Hamiltonian is given by:
\begin{equation}
H_0=\sum_{{\bm n},{\bm p},\xi,\xi'} |{\bm n},\xi\rangle h_{\bm p}^{\xi \xi'} \langle {\bm n}+{\bm p},\xi'| = \sum_{{\bm n},{\bm p}} |{\bm n}\rangle \hat{h}_{\bm p} \langle {\bm n}+{\bm p}|,
\end{equation}
where ${\bm p}$ runs over first, second, etc., neighbors of the origin, $|{\bm n}\rangle $ denotes the one column matrix with the entries $|{\bm n},\xi\rangle$, $\langle {\bm n}|$ represents its dual, and $\hat{h}_{\bm p}$ is the matrix of elements $h_{\bm p}^{\xi \xi'}$. We are interested in the properties of a disordered Hamiltonian $H_\omega=H_0+V_\omega$, and to be specific we consider an onsite random potential:
\begin{equation}
V_\omega = W \sum_{\bm n,\xi} \omega_{{\bm n}\xi} |{\bm n},\xi\rangle \langle {\bm n},\xi|=W \sum_{\bm n}  |{\bm n}\rangle \hat{v}_\omega \langle {\bm n}|
\end{equation}
where $\omega_{{\bm n}\xi}$ are randomly independent amplitudes uniformly distributed in the interval $[-\frac{1}{2},\frac{1}{2}]$.

The twisted boundary conditions method for disordered systems consists basically in considering a large super-cell ${\cal S}$, containing many unit cells of the clean system, specifically the states $|{\bm n},\xi\rangle$, $0\leq n_1, n_2, n_3 \leq {\cal N}-1$, and a random potential is placed inside this super-cell. The super-cell is then periodically repeated in space. We will continue to use $H_\omega$ to denote the resulting approximate Hamiltonian. Since we are dealing with a periodic system, we can construct a dual $k$-space representation via the Bloch transformation. This transformation is given by the isometry ${\cal U}$ from the Hilbert space ${\cal H}$ of the infinitely repeated system into a continuum direct sum of copies of the supercell's Hilbert space ${\cal H}_0$ spanned by $|{\bm n},\xi\rangle$, $0\leq n_1, n_2, n_3 \leq {\cal N}-1$ (${\cal T}$ = 3D torus):
\begin{equation}
\begin{array}{c}
{\cal U}: {\cal H} \rightarrow \bigoplus\limits_{{\bm k}\in {\cal T}} {\cal H}_0, \medskip \\
{\cal U}|{\bm n}+{\bm m}{\cal N},\xi\rangle = \frac{1}{(2\pi)^{3/2}} \bigoplus\limits_{{\bm k}\in {\cal T}}  e^{-i {\bm k}\cdot{\bm m}} |{\bm n},\xi\rangle,
\end{array}
\end{equation}
for any ${\bm n}$ in ${\cal S}$ and arbitrary ${\bm m}$ in ${\bm Z}^d$. Note that any point from ${\bm Z}^d$ can be uniquely written as ${\bm n}+{\bm m}{\cal N}$. The inverse of the isometry is:
\begin{equation}
\begin{array}{c}
{\cal U}^{-1}\left ( \bigoplus\limits_{{\bm k}'\in {\cal T}}\delta_{{\bm k}'{\bm k}}|{\bm n},\xi \rangle \right ) \medskip \\
= \frac{1}{(2\pi)^{3/2}}   \sum\limits_{\bm m\in {\bm Z}^3} \ e^{i {\bm k}\cdot{\bm m}} |{\bm n}+{\bm m}{\cal N},\xi\rangle.
\end{array}
\end{equation}
Note that we wrote the transformation so that all the states inside the supercell get the same phase factor, in which case the $2\pi$-periodicity in the $k$-variables is automatically satisfied. Under this transformation we have:
\begin{equation}\label{QSHBloch}
\begin{array}{c}
{\cal U}H_\omega {\cal U}^{-1}=\bigoplus_{{\bm k}\in {\cal T}} H_\omega({\bm k}),
\end{array}
\end{equation}
where the Bloch Hamiltonians $H_\omega({\bm k}):{\cal H}_0 \rightarrow {\cal H}_0$  are defined by:
\begin{equation}
\begin{array}{c} 
H_\omega({\bm k})=\sum\limits_{ {\bm n} \in {\cal S}}\sum\limits_{\bm p} |{\bm n}\rangle \ \hat{h}'_{{\bm n} {\bm p}}({\bm k}) \ \langle ({\bm n}+{\bm p})\mbox{mod}{\cal N}| \medskip \\
+ W \sum\limits_{{\bm n}\in{\cal S}}  |{\bm n}\rangle \hat{v}_\omega \langle {\bm n}|
\end{array}
\end{equation}
with
\begin{equation}
\hat{h}'_{{\bm n} {\bm p}}({\bm k})=
e^{i\sum\limits_{\alpha=1}^3 k_\alpha(\delta_{n_\alpha+p_\alpha,{\cal N}}-\delta_{n_\alpha+p_\alpha,-1})}\hat{h}_{{\bm p}}. 
\end{equation} 
These are precisely the twisted boundary conditions since the phase factor above occurs only for the lattice points ${\bm n}$ at the boundary. At this point we obtained a family of Bloch Hamiltonians indexed by a point on the 3D torus. The same construction can be achieved by wrapping the supercell ${\cal S}$ into a 3D torus and by threading magnetic fluxes through the 2D sections of this torus. The effect of such magnetic fluxes is captured by the same twisted boundary conditions. 

One should note that the twisted boundary conditions method is defined for a finite volume, whereas one is really interested in the infinite bulk samples. Strictly speaking, one has to take the volume of the supercell to infinity and carefully investigate the stability of the results. In practice, of course, we will have to stop the limit at some point.

\begin{figure}
  \includegraphics[width=5cm]{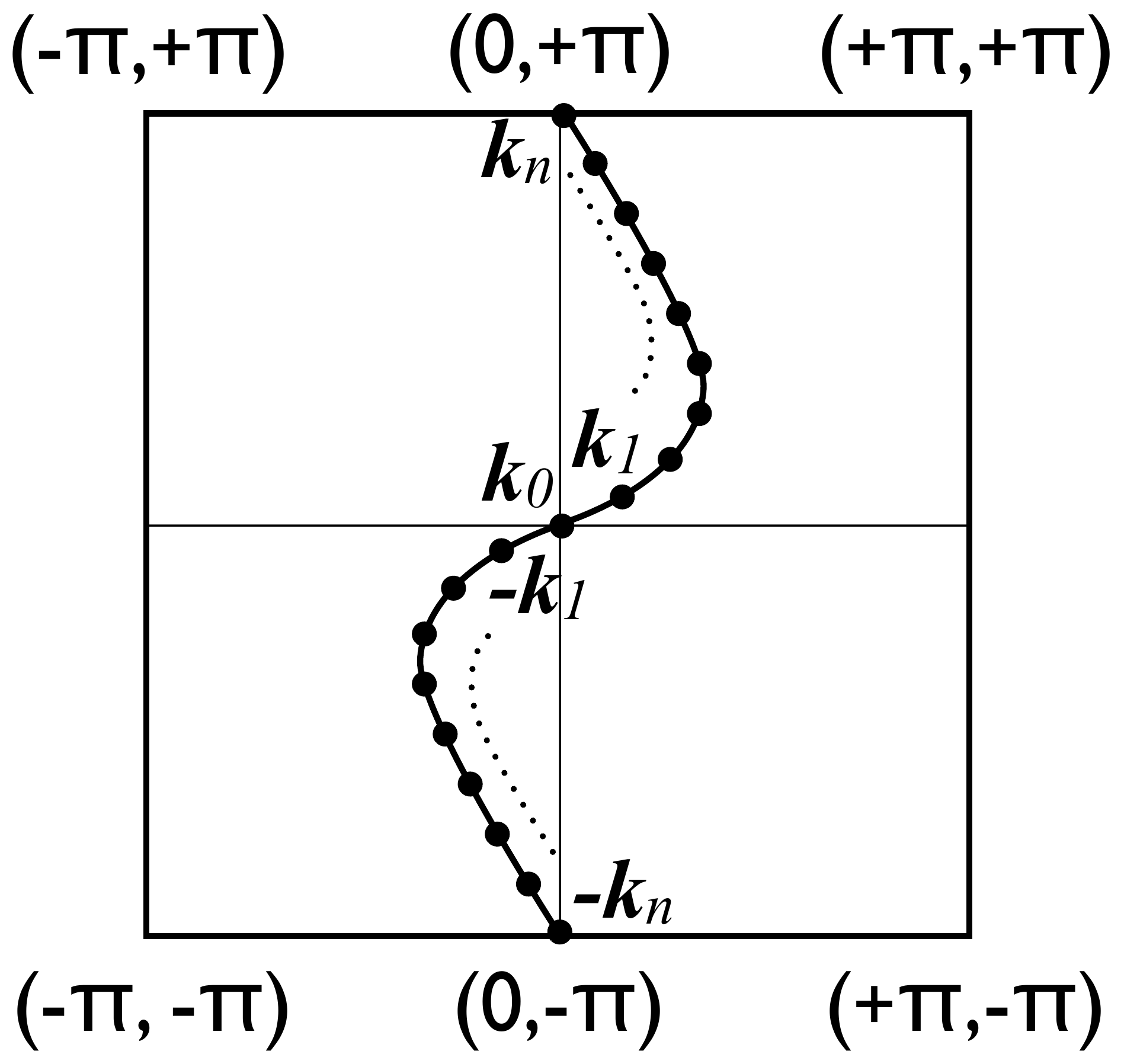}\\
  \caption{Example of a time-reversal invariant path in the Brillouin torus and its discretization.}
 \label{Fig1}
\end{figure}

\section{The ${\bm Z}_2$ invariants}    

We start our discussion of the ${\bm Z}_2$ invariant with a brief review of the formulation given in Ref.~\onlinecite{ProdanPRB2011vy}. The connection between this formulation and the previously existing ones has been exhaustively discussed in this reference and will not be addressed here. Instead, we will give a detailed discussion of the numerical advantages brought in by this new method.

Briefly, the method goes as follows. Let $P_{\bm k}$ denote the projector on to the states of $H_\omega({\bm k})$ below the Fermi level $E_F$, and let $\theta$ denote the time-reversal operation and assume the time-reversal invariance:
\begin{equation}
\theta H_\omega({\bm k}) \theta^{-1} = H_\omega(-{\bm k}).
\end{equation}
One considers a closed, time-reversal invariant path (i.e. a path which is mapped into itself by $\theta$) on the Brillouin torus, 
\begin{equation}
[-\pi,\pi]\ni k \rightarrow {\bm k}(k),
\end{equation} 
parametrized by the variable $k$ (we chose the notation on purpose because in practice this variable will often be $k_z$ for example). Then one integrates the differential equation (with the initial condition $U_{k',k'}=P_{{\bm k}(k')}$): 
\begin{equation}\label{AdTransp}
\begin{array}{c}
i\frac{d}{d k} U_{k,k'} =i [P_{{\bm k}(k)},\partial_{k} P_{{\bm k}(k)}] U_{k,k'}.
\end{array}
\end{equation}
We will use the simplified notation: $P_{{\bm k}(k)}=P_k$. The result of the integration gives the unitary time evolution operator $U_{k,k'}$ corresponding to the process of adiabatically changing the ${\bm k}$-vector along the chosen path in the Brillouin torus. It is assumed that the path starts ($k=-\pi$) and closes ($k=\pi$) at a time-reversal invariant ${\bm k}$-point. Necessarily, the path will cross another time-reversal invariant point at midway $k=0$ (see Fig.~1). Next, one considers arbitrary bases $\{e^0_\alpha\}$ and $\{e^\pi_\alpha\}$ for the occupied spaces at the time-reversal invariant points $k=0$ and $k=\pi$, respectively, and one defines the following matrices:
\begin{equation}
\begin{array}{c}
\hat{U}_{\alpha \beta}=\langle e^\pi_\alpha|U_{\pi,0}|e^0_\beta\rangle, \medskip \\
\hat{\theta}^0_{\alpha \beta}=\langle e^0_\alpha|\theta |e^0_\beta \rangle, \medskip  \\ 
\hat{\theta}^\pi_{\alpha \beta}=\langle e^\pi_\alpha|\theta |e^\pi_\beta \rangle. 
\end{array}
\end{equation}
These matrices satisfy the following fundamental relation:\cite{ProdanPRB2011vy}
\begin{equation}\label{Main}
\frac{\mbox{Pf}\{\hat{\theta}_\pi\}^{-1} \det \{ \hat{U} \} \mbox{Pf}\{ \hat{\theta}_0 \} }{ \sqrt{\det \{U_{\pi,-\pi}\}} }= \pm 1.
\end{equation} 
The left hand side of Eq.~\ref{Main} will be called a pseudo ${\bm Z}_2$ invariant for the following reasons. The left hand side is gauge independent. Indeed, given the transformation properties of the Pfaffians and determinants under the conjugation with unitary matrices, one can easily see that the numerator is independent of the bases $\{e^0_\alpha\}$ and $\{e^\pi_\alpha\}$.\cite{ProdanPRB2011vy} At the denominator, inside the square root, $U_{\pi,-\pi}$ maps the $k=\pm \pi$ occupied space into itself, so at a change of  $\{e^\pi_\alpha\}$ basis we have $U_{\pi,-\pi} \rightarrow E U_{\pi,-\pi} E^{-1}$, with $E$ a unitary matrix, so the determinant remains unchanged. However, the sign in Eq.~\ref{Main} depends on which branch of the square root is used, but once a choice is made the value of the left-hand side cannot be changed by smooth deformations of the Bloch Hamiltonians that keep the insulating gap opened. 

In practice, the adiabatic evolution operators is computed by discretizing the paths and taking the product of projectors onto the occupied spaces at these discrete $k$-points. Since the path is time-reversal invariant, we can choose the discretization points so as ${\bm k}_0$, ${\bm k}_1$, \ldots, ${\bm k}_n$ discretizes the path from $k=0$ to $k=\pi$, while $-{\bm k}_n$, $-{\bm k}_{n-1}$, ...,${\bm k}_0$ discretizes the path from $k=-\pi$ to $k=0$. In this case:
\begin{equation}
U_{\pi,-\pi}=\lim_{n \rightarrow \infty} P_{{\bm k}_n}P_{{\bm k}_{n-1}}\ldots P_{{\bm k}_0} \ldots P_{-{\bm k}_{n-1}}P_{-{\bm k}_n}.
\end{equation}
In practice however, we have to stop limit at some $n=\bar{n}$ and work with an approximation:
\begin{equation}\label{MonFormula}
U_{\pi,-\pi}= P_{{\bm k}_{\bar{n}}}P_{{\bm k}_{\bar{n}-1}}\ldots P_{{\bm k}_0} \ldots P_{-{\bm k}_{\bar{n}-1}}P_{-{\bm k}_{\bar{n}}},
\end{equation}
and similar for $U_{\pi,0}$:
\begin{equation}
U_{\pi,0}= P_{{\bm k}_{\bar{n}}}P_{{\bm k}_{\bar{n}-1}}\ldots P_{{\bm k}_0}.
\end{equation}
We are going to show in the following that the quantization in Eq.~\ref{Main} remains exact even for finite $\bar{n}$'s.

Indeed, using the elementary fact that $\theta P_{{\bm k}_j} \theta^{-1}=P_{-{\bm k}_j}$, we have:
\begin{equation}
U_{\pi,-\pi}=P_{{\bm k}_{\bar{n}}}P_{{\bm k}_{\bar{n}-1}}\ldots P_{{\bm k}_0} \theta P_{{\bm k}_0} \ldots  P_{{\bm k}_{\bar{n}-1}}P_{{\bm k}_{\bar{n}}}\theta^{-1}
\end{equation}
Inserting the identity operator $\sum_\alpha |e^0_\alpha \rangle \langle e^0_\alpha|$  at the appropriate places, we obtain:
\begin{equation}
\begin{array}{c}
\langle e^\pi_\alpha| U_{\pi,-\pi} |e^\pi_\beta \rangle = \langle e^\pi_\alpha|P_{{\bm k}_{\bar{n}}}P_{{\bm k}_{\bar{n}-1}}\ldots P_{{\bm k}_0} |e^0_\delta \rangle \langle e^0_\delta|\theta|e^0_\gamma\rangle \medskip \\
\times \overline{\langle e^0_\gamma|P_{{\bm k}_0} \ldots  P_{{\bm k}_{\bar{n}-1}}P_{{\bm k}_{\bar{n}}}|e^\pi_\xi \rangle \langle e^\pi_\xi| \theta^{-1}|e^\pi_\beta\rangle } \medskip \\
=\langle e^\pi_\alpha|P_{{\bm k}_{\bar{n}}}P_{{\bm k}_{\bar{n}-1}}\ldots P_{{\bm k}_0} |e^0_\delta \rangle (\hat{\theta}_0)_{\delta\gamma} \medskip \\
\times\langle e^\pi_\xi|P_{{\bm k}_{\bar{n}}}P_{{\bm k}_{\bar{n}-1}} \ldots P_{{\bm k}_0}|e^0_\gamma\rangle (\hat{\theta}^{-1}_\pi)_{\xi \beta}.
\end{array}
\end{equation}
Summation over repeating indices was assumed above. At this step, the conclusion is:
\begin{equation}
U_{\pi,-\pi}=\hat{U} \hat{\theta}_0 \hat{U}^T\hat{\theta}_\pi^{-1}.
\end{equation}
Taking the determinant and using the elementary properties of the determinants and pfaffians we obtain:
\begin{equation}
\det \{U_{\pi,-\pi} \} =[\mbox{Pf}\{\hat{\theta}_\pi\}^{-1} \det \{ \hat{U} \} \mbox{Pf}\{ \hat{\theta}_0 \}]^2,
\end{equation}
which is precisely Eq.~\ref{Main}.

The significance of the above conclusion for the numerical calculations is that it allows us to use relatively small number of discretization points when evaluating Eq.~\ref{Main}. One question that could be asked is if the result of such calculation, while indeed quantized, it really equals the result in the $\bar{n}\rightarrow \infty$ limit? To answer this question, we imagine a calculation with a dense number of discretization points and then adiabatically collapsing pairs of adjacent discretization points into a single discretization point. In this way we can adiabatically transform the original computation into a computation with half the number of discretized points. Repeating the same action we can adiabatically reduce the number of discretization points even further, by 4, 8 and so on. Since Eq.~\ref{Main} is quantized, it cannot change its value during such adiabatic deformations, if all the quantities remain well defined. So what can go wrong? In the $\bar{n}\rightarrow \infty$ limit, $U_{\pi,-\pi}$ is a true unitary operator so its determinant is a complex number on the unit circle. For finite $\bar{n}$, $U_{\pi,-\pi}$ is no longer unitary and its determinant moves inside the unit circle. As the number of discretization points is reduced, the determinant moves closer to the origin so there is the possibility that $\det\{U_{\pi,-\pi}\}$ actually becomes equal to zero. At such instance, the calculation breaks down and the quantized value of Eq.~\ref{Main} can change. So the conclusion is that Eq.~\ref{Main} can be indeed evaluated using a relatively small number of discretization points, as long as one makes sure that $\det\{U_{\pi,-\pi}\}$ does not touches the origin. In practice we choose the number of  discretized  points so that $|\det\{U_{\pi,-\pi}\}|\approx 0.5$, which reduces the number of required $k$-points by an order of magnitude in our calculations, when compare with the case when $|\det\{U_{\pi,-\pi}\}|\approx 0.9$.

Eq.~\ref{Main} is fundamental for the formulation of the ${\bm Z}_2$ invariant but it cannot define an invariant by itself. That is because we don't have a canonical way to choose the branch of the square root at the denominator of Eq.~\ref{Main}. However, the important observation is that if one considers a pair of paths, then there is a canonical way to choose the same branch of the square root and genuine ${\bm Z}_2$ invariants can be defined. This has been detailed in Ref.~\onlinecite{ProdanPRB2011vy}. The following lines explain how the procedures were explicitly implemented in our calculations.

For a 3D system, we consider 4 independent time-reversal invariant paths. If ${\cal P}_{k_x,k_y}$ denotes the path along $k_z$ direction that intersects the plane $k_z=0$ at $(k_x,k_y)$, then we choose the following 4 paths:
\begin{equation}\label{Paths}
\begin{array}{l}
{\cal P}_{0,0}: \ (0,0,-\pi) \rightarrow (0,0,\pi) \\ 
{\cal P}_{0,\pi}: \ (0,\pi,-\pi) \rightarrow (0,\pi,\pi)\\
{\cal P}_{\pi,0}: \ (\pi,0,-\pi) \rightarrow (\pi,0,\pi)\\
{\cal P}_{\pi,\pi}: \ (\pi,\pi,-\pi) \rightarrow (\pi,\pi,\pi),
\end{array}
\end{equation}
We interpolated between the paths ${\cal P}_{0,0}$ and ${\cal P}_{0,\pi}$ using the process:
\begin{equation}
[0,\pi] \ni k_y \rightarrow {\cal P}_{0,k_y}.
\end{equation}
By computing the adiabatic evolution 
\begin{equation}
U_{(0,k_y,-\pi)\rightarrow(0,k_y,\pi)}
\end{equation} 
for the path ${\cal P}_{0,k_y}$, we continuously interpolate between the determinants 
\begin{equation}\label{Dets}
\det\{U_{(0,0,-\pi)\rightarrow(0,0,\pi)} \} \ \leftrightarrow \  \det\{U_{(0,\pi,-\pi)\rightarrow(0,\pi,\pi)}\},
\end{equation} 
using the process:
\begin{equation}\label{Interpol}
[0,\pi] \ni k_y \rightarrow \det\{U_{(0,k_y,-\pi)\rightarrow(0,k_y,\pi)}\}.
\end{equation}
This allows us to monitor how the determinant moves on the Riemann surface of the square root function, and to determine the location of $\det\{U_{(0,\pi,-\pi)\rightarrow(0,\pi,\pi)}\}$ relative to the location of $\det\{U_{(0,0,-\pi)\rightarrow(0,0,\pi)} \}$ on the Riemann surface. If $\det\{U_{(0,k_y,-\pi)\rightarrow(0,k_y,\pi)}\}$ crosses the semi-axis $(-\infty,0)$ an odd number of times, then the determinants are located on opposite Riemann sheets and we have to use different branches of the square root, that is, we will have to use $\pm \sqrt{z}$ for one determinant and $\mp \sqrt{z}$ for the other determinant in Eq.~\ref{Dets} when we evaluate the denominator of Eq.~\ref{Main}. If $\det\{U_{(0,k_y,-\pi)\rightarrow(0,k_y,\pi)}\}$ crosses the semi-axis $(-\infty,0)$ an even number of times, then the determinants are located on the same Riemann sheet and we have to use $\pm \sqrt{z}$ for one determinant and same $\pm \sqrt{z}$ for the other determinant. As one can see, there is still a sign ambiguity remaining (originally we had two sign ambiguities) but that becomes irrelevant if we form the product of two pseudo-invariants.
Indeed, the following quantity:
\begin{eqnarray}\label{Xi0}\nonumber
\Xi_{0}=\frac{\mbox{Pf}\{\hat{\theta}_{(0,0,\pi)}\}^{-1} \det \{ \hat{U}_{(0,0,0) \rightarrow (0,0,\pi)} \} \mbox{Pf}\{ \hat{\theta}_{(0,0,0)} \} }{ \sqrt{\det \{U_{(0,0,-\pi)\rightarrow(0,0,\pi)}\}} } \medskip \\
\nonumber \times \frac{\mbox{Pf}\{\hat{\theta}_{(0,\pi,\pi)}\}^{-1} \det \{ \hat{U}_{(0,\pi,0) \rightarrow (0,\pi,\pi)} \} \mbox{Pf}\{ \hat{\theta}_{(0,\pi,0)} \} }{ \sqrt{\det \{U_{(0,\pi,-\pi)\rightarrow(0,\pi,\pi)}\}} }
\end{eqnarray}
is a genuine invariant taking the quantized values $\pm 1$, which are independent of the branch of the square roots used in the calculation, as long as they are chosen consistently using the interpolating procedure described above. We can repeat the same construction for the pair of paths ${\cal P}_{\pi,0}$ and ${\cal P}_{\pi,\pi}$ and define the invariant:
\begin{eqnarray}\label{Xipi}\nonumber
\Xi_{\pi}=\frac{\mbox{Pf}\{\hat{\theta}_{(\pi,0,\pi)}\}^{-1} \det \{ \hat{U}_{(\pi,0,0) \rightarrow (\pi,0,\pi)} \} \mbox{Pf}\{ \hat{\theta}_{(\pi,0,0)} \} }{ \sqrt{\det \{U_{(\pi,0,-\pi)\rightarrow(\pi,0,\pi)}\}} } \medskip \\
\nonumber \times \frac{\mbox{Pf}\{\hat{\theta}_{(\pi,\pi,\pi)}\}^{-1} \det \{ \hat{U}_{(\pi,\pi,0) \rightarrow (\pi,\pi,\pi)} \} \mbox{Pf}\{ \hat{\theta}_{(\pi,\pi,0)} \} }{ \sqrt{\det \{U_{(\pi,\pi,-\pi)\rightarrow(\pi,\pi,\pi)}\}} }.
\end{eqnarray}

The invariants $\Xi_0$ and $\Xi_\pi$ are two of the four independent weak ${\bm Z}_2$ invariants. We can define two more weak invariants by pairing the paths in different ways, but that is not necessary because at this point we can define the strong ${\bm Z}_2$ invariant as:
\begin{equation}
\Xi = \Xi_0 \Xi_\pi
\end{equation}
If we count the strong ${\bm Z}_2$ invariant, then there are only 3 independent weak ${\bm Z}_2$ invariants remaining. We will concentrate entirely on the strong invariant.

We have already discussed the numerical aspects related to computing the ${\bm Z}_2$ pseudo-invariants for each of the four paths of Eq.~\ref{Paths}. There is another important numerical aspect about determining the correct branch of the square roots. One should note that computing the pseudo-invariants involves one dimensional calculations, in the sense that we only need to integrate along the $k_z$ direction and not on a surface as it is the case when applying, for example, the popular algorithm of Fukui et al from Ref.~\onlinecite{FukuiJPSJ2007ny}. However, we still have to perform the interpolation along the $k_y$ direction, so the calculations become 2-dimensional. The key observation is that the number of $k_y$ points required by a successful interpolation is usually an order of magnitude smaller than the number of $k_z$ points needed in the computation of the pseudo-invariants. This is because all we need is to determine how the determinants wind around the origin during the interpolation and to trace these paths one can indeed use a relatively small number of $k_y$ points. Therefore, our algorithm, while not strictly 1-dimensional, it can be regarded as quasi-one dimensional. 

To summarize, the application to the disordered system was possible because of the following numerical advantages of the present algorithm:
\begin{enumerate}
\item  The algorithm is gauge independent. Finding a smooth gauge for a unit cell containing thousands of quantum states would have been practically impossible.
\item  The quantization of the pseudo-invariants remain exact when the paths are discretized, allowing a drastic reduction of the number of the discretization points.
\item The interpolation between the different time-reversal invariant paths can be accomplished with a small number of k-points, transforming the algorithm into a quasi-one dimensional one.
\end{enumerate}

\section{The model}

\begin{figure}
  \includegraphics[width=8.6cm]{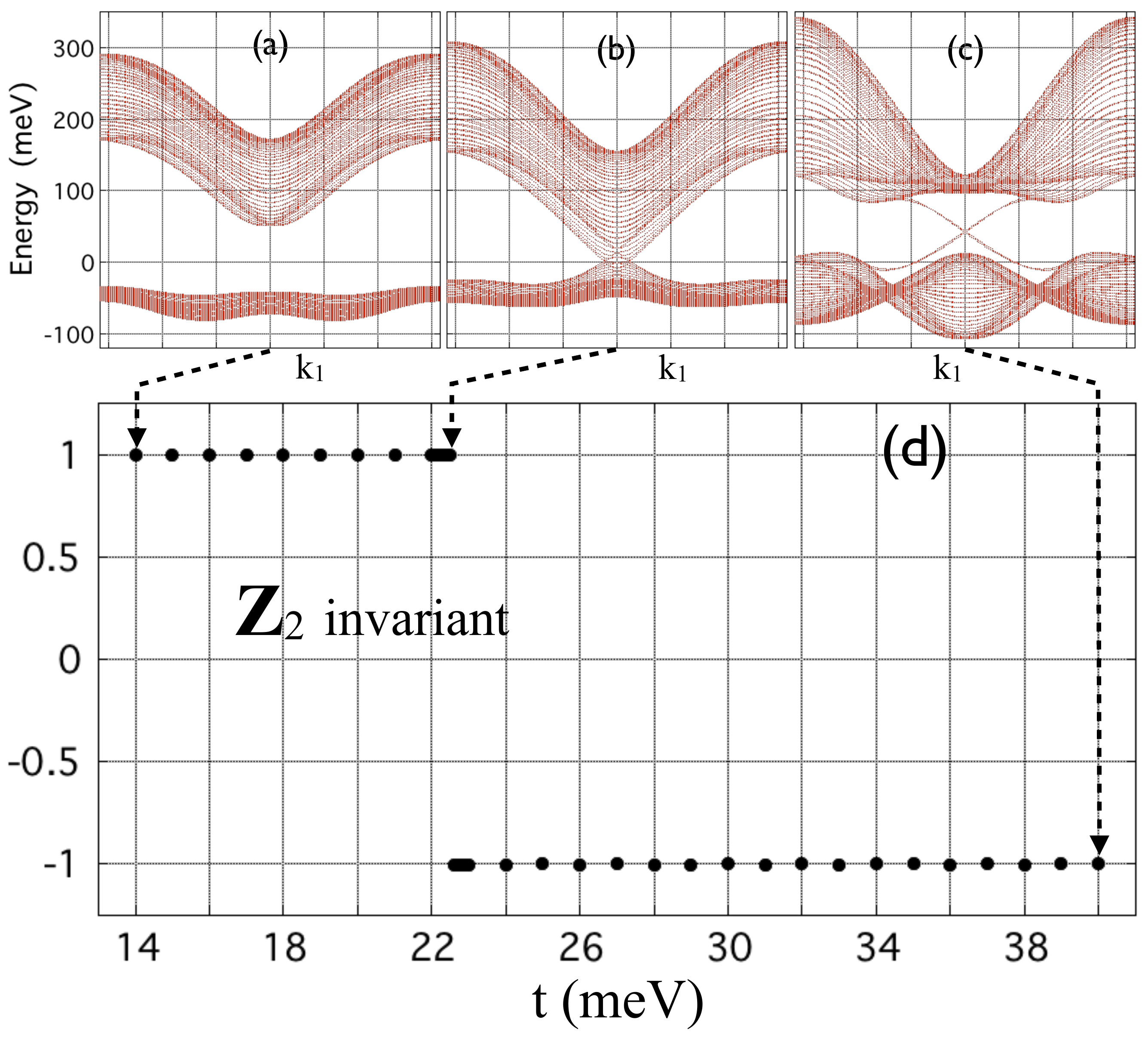}\\
  \caption{(a-c) The band structure of the model for a slab configuration $0<n_3<30$, plotted as function of $k_1$ with $k_2$ fixed at $k_2=0$. The hopping parameter $t$ takes the values $t=14$ meV in panel (a), $t=22.6$ meV in panel (b) and $t=40$ meV in panel (c). Panel (d) reports a calculation of the strong ${\bm Z}_2$ invariant as $t$ was varied from 14 to 40 meV.}
 \label{Fig2}
\end{figure}

The model used in our numerical simulations is an effective lattice Hamiltonian fitted to the topological material Bi$_2$Se$_3$. The starting point is a Hamiltonian $H_0$ which, in the clean limit, can accurately describe the empirical energy band spectrum around the insulating gap. This $H_0$ was used in the previous studies of disordered Bi$_2$Se$_3$ in Refs.~\onlinecite{GuoPRB2010fu,GuoPRL2010vb}. The $H_0$ has inversion symmetry and, since we want to exemplify the algorithms for systems without such symmetry, we will include an additional term in the Hamiltonian that strongly breaks the inversion symmetry. This term can be thought as the effect of a mechanical strain applied along the $z$ axis.

In the momentum space: 
\begin{equation}\label{H3D}
H_0({\bm k})=d_4({\bm k}) +
 \left(
 \begin{array}{cccc}
 d_0({\bm k})&d_3({\bm k})&0&d_{-}({\bm k})\\
 d_z({\bm k})&-d_0({\bm k})&d_{-}({\bm k})&0\\
 0&d_{+}({\bm k})&d_0({\bm k})&-d_3({\bm k})\\
 d_{+}({\bm k})&0&-d_3({\bm k})&-d_0({\bm k})
 \end{array}
 \right)
\end{equation}
where
\begin{equation}
\begin{array}{c}
d_0({\bm
k})=\epsilon-2t\sum_i \cos k_i, \medskip \\
d_i({\bm k})=-2\lambda \sin k_i, \ \ i=1,2,3 \medskip \\
d_4({\bm k})=2\gamma \left (3-\sum_i \cos k_i \right)
\end{array}
\end{equation}
and
\begin{equation}
d_{\pm}({\bm k})=d_1({\bm k})\pm i d_2({\bm k}).
\end{equation}
The added term that preserves the time-reversal symmetry but breaks the inversion symmetry is:
  \begin{equation}\label{H3D3}
 V_{\cal I}=R\left (
 \begin{array}{cccc}
 0 & 0 & 0 & e^{-ik_3} \\
 0 & 0 & -e^{ik_3} & 0 \\
 0 & -e^{-ik_3} & 0 & 0 \\
 e^{ik_3} & 0 & 0 & 0
 \end{array}
 \right )
\end{equation}
The following parameters will be fixed at these values throughout the paper: $\epsilon=134$ meV, $\lambda=30$ meV, $\gamma=16$ meV, $R=15$ meV. We will use $t=40$ meV for the topological insulator and $t=14$ meV for the trivial insulator (the two values lead to comparable insulating gaps). The insulating gap in our study is larger than the empirical insulating gap of the Bi$_2$Se$_3$ material, and the reason we chose to proceed this way was to be able to better showcase the behavior of the strong ${\bm Z}_2$ invariant in the presence of disorder (the insulating and the mobility gaps would have closed too fast if the gap was fixed at the empirical value). This modification does not break the bridge with the experimental reality because it is known that a mechanical strain may increase the insulating gap of the material.

The real space representation of the translational invariant Hamiltonian can be constructed on cubic lattice where each vertex ${\bm n}$ carries four quantum states $|{\bm n},\alpha,\sigma\rangle $. Here, $\alpha=\pm 1$ (= isospin) labels the $s$ or the $p$ angular momentum character of a state the bands and $\sigma=\pm 1$ the spin up and down configurations. On the Hilbert space spanned by $|{\bm n},\alpha,\sigma\rangle$, we define $d_{i,j,k}$, $\hat{\sigma}$, $\hat{\alpha}$, $r_{\alpha}$ and $r_{\sigma}$ as the translation, spin, isospin and flipping operators as follows:
\begin{equation}
\begin{array}{l}
\hat{d}_{i,j,k} |n_1,n_2,n_3, \alpha,\sigma \rangle = |n_1+i,n_2+j,n_3+k,\alpha,\sigma \rangle \medskip\\
\hat{\sigma} | \bm n, \alpha, \sigma \rangle = \sigma | \bm n, \alpha, \sigma \rangle ,  \quad \hat{\alpha} | \bm n, \alpha, \sigma \rangle= \alpha    | \bm n, \alpha, \sigma \rangle r \medskip\\
 r_{\sigma} | \bm n, \alpha, \sigma \rangle = | \bm n, \alpha, -\sigma \rangle, \quad r_{\alpha} | \bm n, \alpha, \sigma \rangle = | \bm n, -\alpha, \sigma \rangle

\end{array}
\end{equation}
Then, the real space representation of $H_0$ takes the form
\begin{equation}
\begin{array}{c}
H_0= \epsilon \hat{\alpha} + 6 \gamma+ \lambda \displaystyle\sum_{s=\pm 1} s\hat{d}_{0,s,0}r_{\alpha} \hat{\sigma}  r_{\sigma}\medskip \\
 +i \displaystyle\lambda \sum_{s=\pm1}s ( \hat{d}_{s,0,0} r_{\alpha} r_{\sigma}+  \hat{d}_{0,0,s}r_{\alpha} \hat{\sigma}) \medskip \\
 - t (\hat{\alpha}+\gamma) \displaystyle\sum_{s=\pm1}( \hat{d}_{s,0,0}+ \hat{d}_{0,s,0} +  \hat{d}_{0,0,s}  ) 
\end{array}
\end{equation}
The term breaking the inversion symmetry takes the form:
\begin{equation}
\begin{array}{c} 
V_{\cal I}= \frac{R}{2}\hat{\sigma}(\hat{\alpha}-1)(\hat{d}_{0,0,1}r_{\alpha} - \hat{d}_{0,0,-1})r_{\sigma}.
\end{array}
\end{equation}

As it is now well established, the topological properties of the clean model are revealed when restricting the total Hamiltonian $H_0+V_{\cal I}$ on a slab: $0<n_3<N$, where $N$ is taken large enough so that the tunneling between the two surfaces of the slab is negligible.  The slab Hamiltonian takes the form:

\begin{equation}
\begin{array}{c}
H_0(k_1,k_2)=-t  \displaystyle\sum_{s=\pm 1} \hat{d}_{00s} \hat{\alpha} + i \lambda  \displaystyle\sum_{s=\pm 1} s\hat{d}_{00s} r_{\alpha}
\hat{\sigma}\medskip \\
-\gamma  \displaystyle\sum_{s=\pm 1} (\hat{d}_{00s}) +  d'_4(k_1,k_2)+d'_0(k_1,k_2) \hat{\alpha} \medskip\\
 + d_1(k_1) r_{\alpha} r_{\sigma} -id_2(k_2) r_{\alpha}\hat{\sigma}r_{\sigma} 
\end{array}
\end{equation}
where  $d'_4(k_1,k_2)= 2 \gamma \big (3- \sum_{1,2}\cos(k_i) \big) $ and $d'_0(k_1,k_2)=\epsilon-2 t\sum_{1,2}\cos(k_i) $.  For such configuration, the parallel component to the surfaces of the momentum is conserved, so one can plot the energy spectrum as function of $k_1$ and $k_2$.  In Fig.~\ref{Fig2}(a-c) we show sections of such plots, by holding $k_2$ at $k_2=0$, for three different values of $t$. The dense band spectra seen in all three plots correspond to the bulk and one can see a bulk energy gap in panels (a) and (c). The bulk gap is closed in panel (b) and that marks the transition from the trivial to the topological insulator. Indeed, in panel (c) one can observe chiral bands connecting the valence and the conduction bands, and in panel (a) these bands are missing entirely. The chiral bands in panel (c), if plotted as function of $k_1$ and $k_2$, will give rise to a Dirac cone. The transition point between the phases is at $t=22.6$ meV. 

A straightforward test of the algorithm described in the previous section consists of computing the strong ${\bm Z}_2$ invariant for the clean system as function of parameter $t$, and comparing  the output with the appearance or disappearance of the surface states in the slab calculations in Fig.~\ref{Fig1}. Fig.~\ref{Fig1}(d) reports these calculations and indeed both the ${\bm Z}_2$ invariant and the slab calculations predict a trivial insulator for $t<22.6$ meV and a topological insulator for $t$ above this value. The strong ${\bm Z}_2$ invariant was computed using the twisted boundary conditions on a 4$\times$4$\times$4 lattice. The size of the lattice is irrelevant for the clean systems, and we just chose a convenient lattice size in order to test the twisted boundary conditions method. We used X number of $k_z$ points to compute the pseudo-invariants, in which case $|\det\{U_{\pi,-\pi}\}|\approx ZZZ$, and Y number of $k_y$ points to perform the interpolations.

\section{Effect of Disorder: Level Statistics Analysis}

\begin{figure}
  \includegraphics[width=8.6cm]{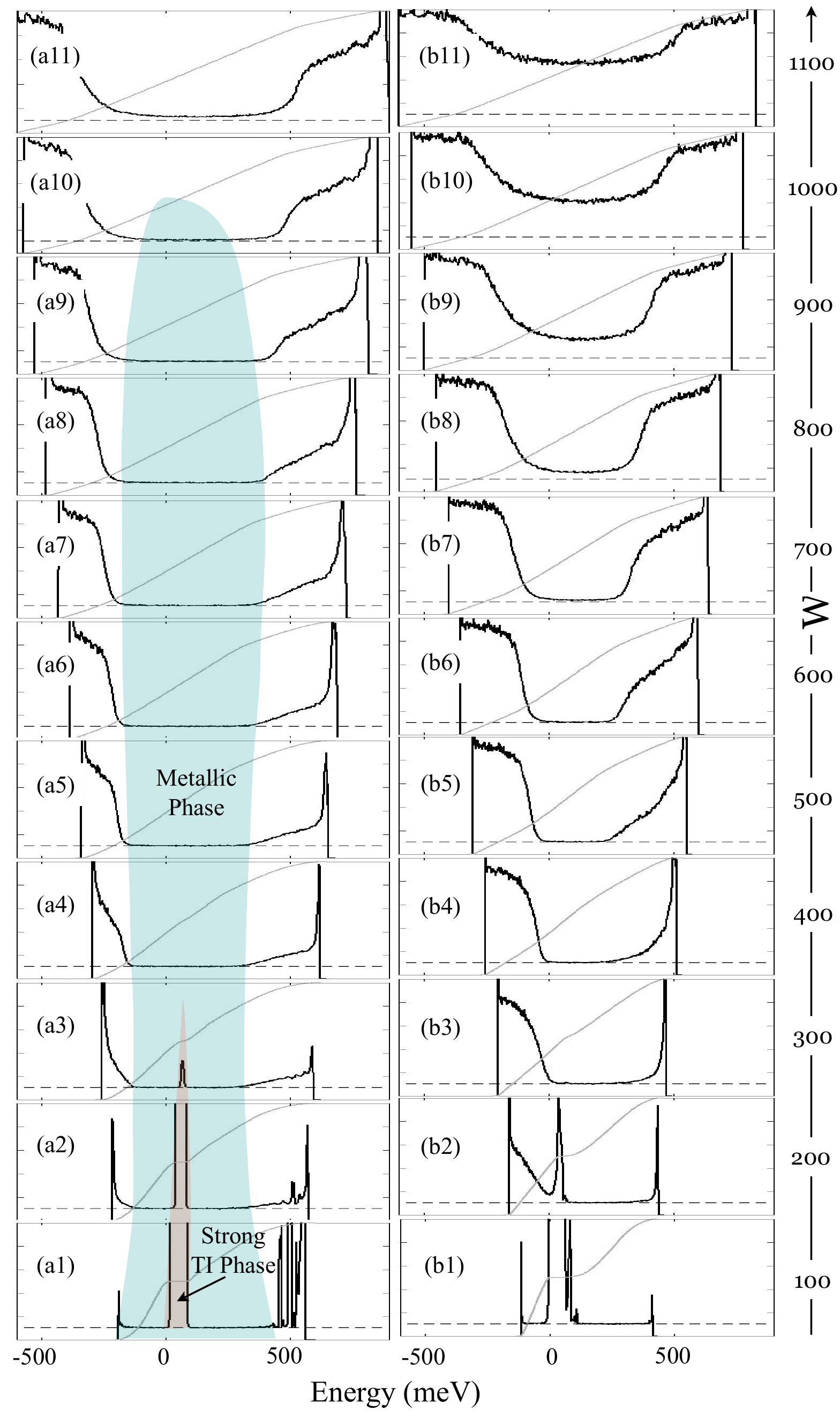}\\
  \caption{Level statistics analysis for (left column) the topological insulator $t=40$ meV and (right column) the trivial insulator $t=14$ meV. Each panel displays the variance of the level spacings ensembles as function of the energy where the level spacings were collected. The gray lines in each panels represent the integrated density of states (IDOS), which can be used to assess the evolution of the spectral gap, corresponding to the flat IDOS, and especially to determine when the gap is closing and becoming completely filled with localized states. The horizontal dash lines mark the value 0.104, the variance of the GSE ensemble. The vertical range in each panel goes from 0 to 1. The shaded regions represent the emerging phase diagram of the topological model.}
 \label{Fig3}
\end{figure}

The total Hamiltonian will include a random potential:
\begin{equation}
H=H_0+V_{\cal I}+V_\omega,
\end{equation}
where $V_\omega$ is a non-magnetic random potential:
\begin{equation}
V_\omega = W\sum_{{\bm n,\alpha,\sigma}}\omega_{{\bm n}\alpha}|{\bm n},\alpha,\sigma \rangle \langle {\bm n},\alpha,\sigma|
\end{equation}
with $\omega_{{\bm n}\alpha}$ random entries uniformly distributed in the interval $[-\frac{1}{2},\frac{1}{2}]$.

 For level statistics analysis, we diagonalized the disordered Hamiltonian on a 14$\times$14$\times$14 lattice with periodic boundary conditions and for 500 random disorder configurations. We sampled the energy spectrum at 100 equally spaced energy levels E. For each such $E$, we identified, for each disorder configuration, the unique energy levels $E_i$ and $E_{i+1}$ satisfying: $E_i<E<E_{i+1}$, and we recorded the level spacings: $\Delta E=E_{i+j+1}- E_{i+j}$, letting $j$ take consecutive values between $-5$ and $5$. Note that  $j$ indexes the levels and that each level is doubly degenerate. In this way, we have generated ensembles containing 5500 level spacings for each energy E. Fig.~\ref{Fig3} reports the variance $\langle s^2\rangle/\langle s \rangle^2-1$ of these ensembles as function of energy $E$ and disorder strength $W$. It also reports the integrated density of states (IDOS), which counts the number of eigenvalues below an energy $E$ and normalizes this number by the dimension of the Hilbert space. When plotted as function of $E$, the IDOS remains flat in the spectral gaps, so it is a useful and effective tool for identifying the spectral gaps in the energy spectrum. We will be particularly interested to see when the insulating gap is closing.
 
The level spacings follow a Poisson distribution when $E$ is in the localized spectrum and the localization length is smaller than the size of the system. The Poisson distribution has a variance equal to 1. In the spectral regions where the localization length exceeds the size of the system, the statistics of the level spacings coincides with that of a random Gaussian Symplectic Ensemble (GSE):\cite{EfetovZhTeorFiz1982wo,EfetovBook1997vn} $P_{\mbox{\tiny{GSE}}}(s)=\frac{2^{18}}{3^6 \pi^3}s^4e^{-\frac{64}{9\pi}s^2}$. The variance of this distribution is 0.104. One can study the trends as the system size is increased, and if the size of the system reached a limit where the variance is seen to stabilize (which we have verified that it does), the level spacing analysis can be used quite effectively to identify the regions of localized and de-localized spectrum. This will be done after we discuss the qualitative behavior of the energy spectrum in response to disorder.

In 3 dimensions, extended states can exist even in trivial disordered models. The qualitative behavior of the spectrum in trivial models is as follows. Usually, the edges of the bands starts to localize the moment the disorder is turned on (for systems displaying large variations in the density of states, additional patches of localized spectrum can occur deep inside the band). At moderate disorder, extended states still survive deep within the bands. As such, there are usually two mobility edges forming per band, flanking the region of extended states, and these mobility edges moves towards each other when the disorder is increased until they merge and disappear. At that point, all the states become localized, as it should be the case at large disorder.\cite{Aizenmann1993uf}

In a topological model, the behavior of the spectrum is markedly different. The edges of the bands are still the first parts of the spectrum to become localized and the extended states are still located in the middle of the bands. But, if a band carries a nontrivial topological number that is robust against disorder, the two mobility edges flanking the extend states in a band cannot merge and disappear like in the trivial case because that will lead to a sudden change of the topological number carried by the bands, from a nontrivial to a trivial value. So what happens when increasing the disorder? The energy spectrum will eventually become entirely localized as the disorder is being steadily increased,\cite{Aizenmann1993uf} and the only way this can happen is through a scenario where  the bands carrying topological numbers collide with each other and in the process they neutralize their topological numbers. This leads to one of the hallmarks of the topological models where, when increasing the disorder, the spectral regions of extended states are seen to drift towards each other until they merge and disappear, usually at very large disorder strengths. The levitation of the Chern-number carrying extended states in the Integer Quantum Hall Effect is well known from the works of Halperin and Laughlin,\cite{HalperinPRB1982er,LaughlinPRL1984mc} and the pair annihilations of the topological states in lattice models of IQHE was discussed in Refs.~\onlinecite{RochePRB1999te,YangPRB1999de}. The levitation and pair annihilation picture was instrumental for the understanding of the global phase diagram of IQHE,\cite{KivelsonPRB1992bv} and that will also be the case for our study.

\begin{figure}
  \includegraphics[width=5.5cm]{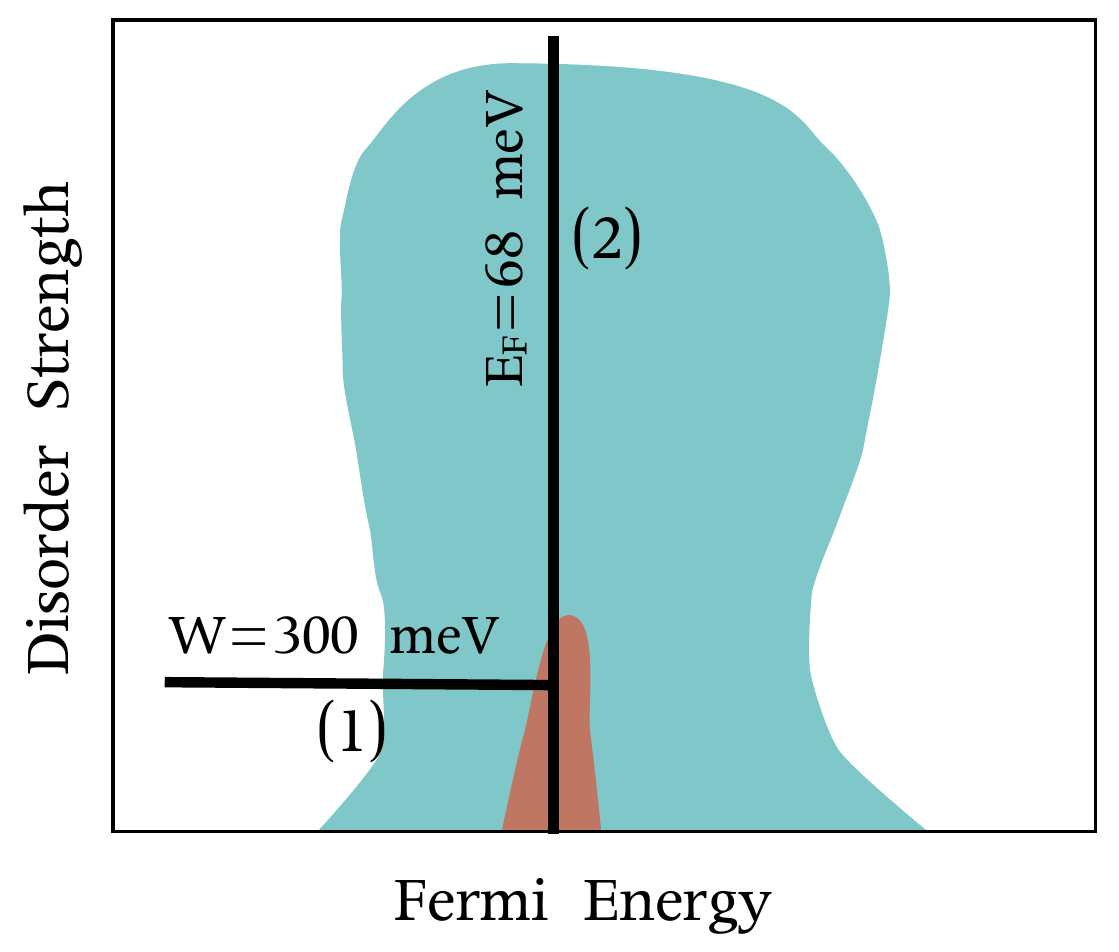}\\
  \caption{Illustration of the phase diagram of the model as derived in Fig.~\ref{Fig3}, and the two paths used in the mapping of the ${\bm Z}_2$ invariant.}
 \label{Fig4}
\end{figure}

In Fig.~\ref{Fig3} we report the variance of the level spacings ensembles collected at various energies and for increasing disorder strengths. We do not show here the actual histograms of the level spacings because of the large volume of data already contained in this figure. However, the level statistics have been exhaustively researched for topological models,\cite{Prodan2010ew,ProdanJPhysA2011xk,Shulman2010cy,Chua2011ci} and the correlation between the histograms and the value of the variance has been already firmly established.\footnote{We thank Andrei Bernevig for suggesting the variance as a good quantity to look at.} Panels (a1)-(a11) refer to the topological case where $t=40$ meV and panels (b1)-(b11) refer to the trivial case where $t=14$ meV. These two values were chosen such that the insulating energy gaps are practically the same (see Fig.~\ref{Fig1}). 

Examining the panels in Fig.~\ref{Fig3}, one can observe energy regions where the variance is large (and becomes unity at large disorder) but also energy regions where the variance remains pinned at the 0.104 value. These later regions will be identified with the spectral regions of extended states, while the former ones with the spectral regions of localized states. In panels (a1)-(a11) we can clearly see the two extended states regions drifting and merging with each other as the disorder is increased. The extended states survive even at extreme values of disorder $W=1000$ meV; this value is about twice the width of the entire clean energy spectrum. No such behavior is observed for the trivial case in panels (b1)-(b11), where the valance band is seen to become entirely localized already at $W$'s as small as  200 meV, and the whole spectrum becomes localized before W reaches 700 meV.

Based on the data presented in Fig.~3(a1)-(a11), we can draw the phase diagram of the topological model in the $(W,E_F)$ plane with quite accurate precision. It consists of a strong topological insulating phase surrounded by a metallic phase, which at its turn is surrounded by a trivial insulating phase (see Fig.~\ref{Fig3}). One could be inclined, by just looking at this 2D phase diagram, to call this later phase the Anderson insulating phase rather than the trivial insulating phase, but this is not correct because if we consider a third dimension to the phase diagram along the parameter $t$, one will easily see that this phase is connected to the trivial insulating phase, such as for example the $t=14$ meV and $w=0$ case shown in Fig.~2. As such, this phase should be called simply the trivial insulating phase. Now, examining the integrated density of states, we can see that the spectral gap is already closed at $W=300$ meV but the topological phase extends beyond this $W$ value. This phase diagram will be re-confirmed by a direct mapping of the ${\bm Z}_2$ invariant. 

\begin{figure*}
  \includegraphics[width=16cm]{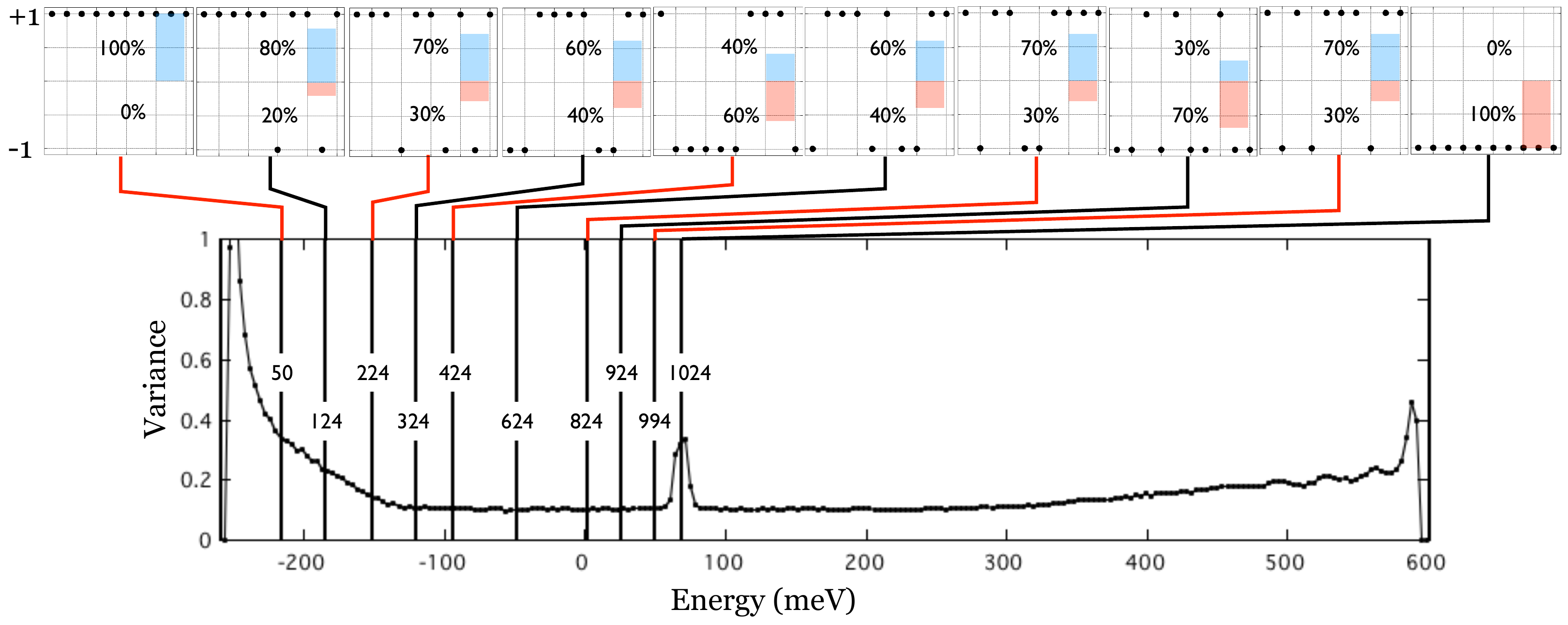}\\
  \caption{The upper panels show the ${\bm Z}_2$ invariant computed along the path (1) of Fig.~\ref{Fig4}, on a 8$\times$8$\times$8 unit cells lattice via twisted boundary conditions. The dimension of the occupied space was slowly reduced from 1024 to 124, as indicated in the figure. Each ${\bm Z}_2$ the calculation was repeated for 10 random disorder configurations and the output is shown by the full dots, exactly how it occur in the actual calculation. The percentages of the ${\bm Z}_2=\pm 1$ occurrences is displayed in each panel. The lower panel shows the variance of the level spacings for $W=300$ meV, and the averaged Fermi levels (see the vertical lines) corresponding to each ${\bm Z}_2$ calculation.}
 \label{Fig5}
\end{figure*}

\section{Maps of the ${\bm Z}_2$ invariant}

The ${\bm Z}_2$ invariant will be mapped along the two paths shown in Fig.~\ref{Fig4}. Due to the extreme computational costs of such calculations, we had to settle for a somewhat smaller lattice size of 8$\times$8$\times$8 (but same as the largest lattice size used in Ref.~\onlinecite{GuoPRB2010fu}). We have used 400 $k$-points in the $k_z$ direction to compute the monodromies, and 25 $k$-points in the $k_y$ direction for the interpolation. As such, each ${\bm Z}_2$ invariant computation requires 20,000 exact diagonalizations of the disordered Hamiltonian. A number of 10 disorder configurations were considered for each point chosen along the paths shown in Fig.~\ref{Fig4}.

There is one important numerical aspect that we must acknowledge, which is the computation of the pfaffian of the time-reversal operator at the time-reversal invariant $k$-points. This became an issue for us because the dimensions of the matrices are very large. We have successfully used the fortran routine PfaffianH freely provided by the authors of Ref.~\onlinecite{BallesteroCPC2011cy}, which computes the pfaffian of a general complex a skew-symmetric matrix using the Householder transformations.

Fig.~\ref{Fig5} reports the map of the ${\bm Z}_2$ invariant along the path (1). The calculations were performed with a fixed number of occupied states rather than a fixed Fermi level. By doing so, we ensured that all the projectors in the monodromy formula Eq.~\ref{MonFormula} have the same dimension but in this case the Fermi level displays small fluctuations which disappear in the thermodynamic limit. The dimension of the occupied states was slowly reduced from 1024 (half-filled) to 50, as illustrated in Fig.~\ref{Fig5}, and for each dimension we have computed the average Fermi level, defined as half between the last occupied and lowest un-occupied states. The averaged Fermi levels are shown as vertical lines, over-imposed on the variance plot at $W=300$ meV. As one can see, the Fermi levels sample the entire spectrum below the gap.

We want to point out again that the spectral gap is already closed at $W=300$ meV but, according to the level statistics analysis, there is still a mobility gap opened. We have verified this statement by direct check of the eigenvalue files. Also, the integrated density of states shows an inflection point rather than a plateau. When the Fermi level was inside this mobility gap, we found absolute no fluctuations in the ${\bm Z}_2$ invariant, which turned out to be $-1$ for all 10 random configurations. As the Fermi level is lowered, it enters the region of extended states and here we observed large fluctuations. As we already discussed, even in this regime the ${\bm Z}_2$ continue to take quantized $\pm 1$ but there is no way to tell which one will be so the output fluctuates between the two allowed values. This remains the case for as long as the Fermi level is in the region of extended states and, as soon as the Fermi level emerges back into the region of localized states, the ${\bm Z}_2$ invariant is seen to take a non-fluctuating quantized value of $+1$.

A similar behavior is observed when considering the path (2) of Fig.~\ref{Fig4} for which  the calculations are reported in Fig.~6. While increasing the disorder strength, the ${\bm Z}_2$ invariant is seen to take the quantized and non-fluctuating value of $-1$ until the mobility gap closes. From there on, the values fluctuate between $\pm 1$, and the ${\bm Z}_2$ stabilizes once again when the path enters the trivial insulating state where it assumes the value $+1$.

\begin{figure}
  \includegraphics[width=8.6cm]{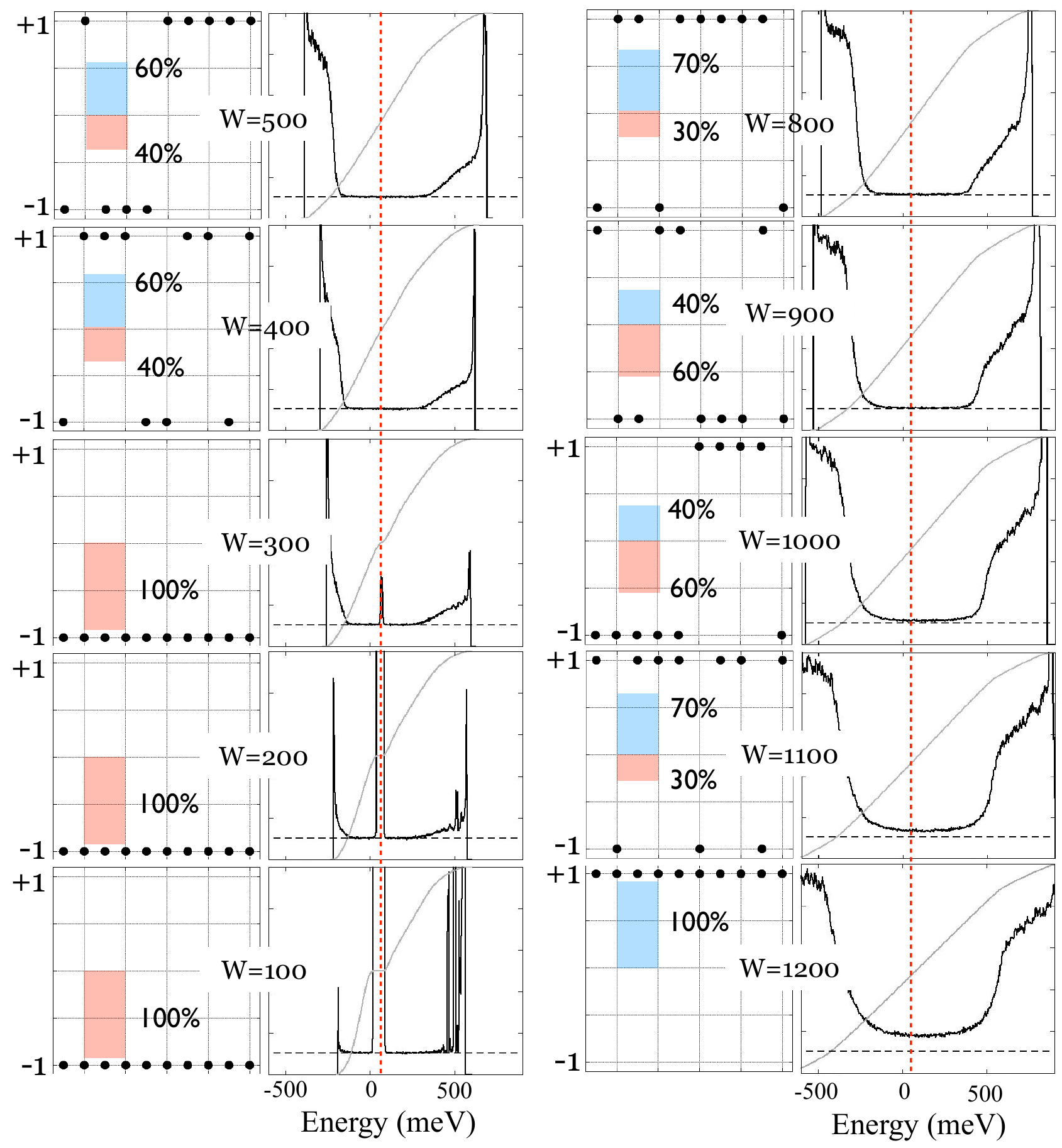}\\
  \caption{This figure reports the results of a computation of the strong ${\bm Z}_2$ invariant along the path (2) of Fig.~\ref{Fig4}, completed on a 8$\times$8$\times$8 unit cells lattice via twisted boundary conditions. The disorder strength was increased from $W=100$ to 1200 meV, as indicated in the figure. Each ${\bm Z}_2$ the calculation was repeated for 10 random disorder configurations and the output is shown by the full dots, exactly how it occur in the actual calculation. The percentages of the ${\bm Z}_2=\pm 1$ occurrences are displayed in each panel. The accompanying panels show the variance of the level spacings at the corresponding $W$'s, from where one can determine when the Fermi level is in a region of localized/delocalized spectrum. The Fermi level, represented by the dotted vertical line, was kept at 68 meV during these calculations.}
 \label{Fig6}
\end{figure}

\section{Conclusions}

In conclusion, a previously introduced gauge-independent formulation of the strong ${\bm Z}_2$ invariant was found to bring significant numerical advantages, allowing direct computations of the invariant for large super-cells with twisted boundary conditions. The resulting numerical algorithm was applied to a disordered model of Bi$_2$Se$_3$ topological material and maps of the strong invariant were given as function of either Fermi level or disorder strength. The behavior of the strong ${\bm Z}_2$ invariant seen in our numerical calculations is exactly what one will expect if this invariant was indeed robust to disorder. Specifically, we observed the strong ${\bm Z}_2$ invariant taking quantized and non-fluctuating values whenever the Fermi level was in an energy region of localized states, and fluctuating values (between the only two possible values of $\pm 1$) whenever the Fermi level was in an energy region of delocalized states. The fact that our numerical maps of the strong ${\bm Z}_2$ invariant were in good agreement with the phase diagram constructed from the level statistics analysis leaves very little doubt that the strong topological phase survives beyond the point where the spectral gap closes, and that it extends all the way to the point where the mobility gap closes.

Our algorithm, combined with accurate tight binding models that can be developed for any material via either first principles calculations or by simple empirical means,\cite{LiuPRB2010xf} can provide accurate quantitative simulations of the real experimental samples. We want to point out that, recently, the twisted boundary conditions were successfully used to compute the Chern invariant of an interacting 2-dimensional fractional Chern insulator.\cite{ShengNatComm2011vy} Since the algorithm for computing the ${\bm Z}_2$ invariants is less demanding than the algorithm for the Chern invariant, we have high hopes that we will soon be able to map the ${\bm Z}_2$ invariants in the presence of electron interaction for accurate complex models of topological materials. Both disorder and electron interactions are expected to strongly influence the phase diagram of a topological material. The topological/non-topological state of a sample can be probed by looking at the extended/localized character of the surface states via transport simulations on quasi-one dimensional bars (like it was done in Ref.~\onlinecite{GuoPRL2010vb}), but here one has to be careful with the boundary conditions and, in addition, the extended surface states can exist in non-topological samples. Therefore, we believe that the direct computation of the strong ${\bm Z}_2$ invariant will be a valuable complement to these aforementioned methods.

The data generated by our study can be of interest for experimentalists. So far, all topological materials fabricated in the labs display metallic bulk properties, a feature that was attributed to the imperfections of the materials. Our study revealed the interesting fact that, due to the very topological nature of the materials, the disorder pulls the valance and the conduction mobility edges closer to each other. In fact, within our tight-binding model for  Be$_2$Se$_3$, we saw a rapid reduction of the mobility gap with disorder and the closer of the mobility gap when the disorder strength reached about 350 meV. This suggests that the topological materials have to be much ``cleaner" than their trivial counterparts in order to see an insulating bulk phase.

\section{Acknowledgements} This research was also supported by the U.S. NSF grants DMS-1066045 and DMR-1056168.


%

\end{document}